\documentclass[twocolumn,final,aps,prb,showpacs,superscriptaddress,amsmath,amssymb,amsfonts,floatfix]{revtex4-2}

\usepackage[utf8]{inputenc}
\usepackage[T1]{fontenc}        % Important for anything that uses more than ASCII
\usepackage{lmodern}            % Beautiful fonts especially for pdfs
\usepackage{graphicx}
\usepackage{multirow}
\usepackage{epsfig}
\usepackage{url}
\usepackage{xcolor}
 \usepackage{ulem}

\usepackage[colorlinks,breaklinks,bookmarks=true,citecolor=blue,linkcolor=blue,urlcolor=blue]{hyperref}
\usepackage{tabularx}
\usepackage{sidecap}
\usepackage{float}

\usepackage{xr-hyper} % Required for cross-referencing
\usepackage{hyperref}
\usepackage{orcidlink}

\usepackage{array}
\allowdisplaybreaks

\usepackage[capitalize]{cleveref}
\usepackage{textgreek}

\crefname{figure}{Fig.}{Figs.}

\newcommand{\rep}[1]{{\color{black} #1}} 
\newcommand{\kh}[1]{{\color{black}#1}}

\newcommand{\fb}[1]{{\color{black} #1}}
\newcommand{\mk}[1]{{\textcolor{black}{#1}}}

\newcommand{\ak}[1]{{\color{black} #1}}

\begin{document}
\title{Entanglement in the pseudogap regime of
  %the Hubbard model and
  cuprate superconductors}

\author{Frederic Bippus \orcidlink{0009-0006-4316-6547}}
\affiliation{%
    Institute of Solid State Physics, TU Wien, 1040 Vienna, Austria}%

\author{Juraj Krsnik \orcidlink{0000-0002-4357-2629}}
\affiliation{%
    Institute of Solid State Physics, TU Wien, 1040 Vienna, Austria}%
\affiliation{%
    Department for Research of Materials under Extreme Conditions, Institute of Physics, HR-10000 Zagreb, Croatia}%
\author{Motoharu Kitatani \orcidlink{0000-0003-0746-6455}}
\affiliation{%
    Department of Material Science, University of Hyogo, Ako, Hyogo 678-1297, Japan}%
%\affiliation{%
%    RIKEN Center for Emergent Matter Sciences (CEMS), Wako, Saitama 351-0198, Japan}%

\author{Luka Ak\v{s}amovi\'{c}}
\affiliation{%
    Institute of Solid State Physics, TU Wien, 1040 Vienna, Austria}%
    
\author{Anna Kauch\,\orcidlink{0000-0002-7669-0090}}
\affiliation{%
    Institute of Solid State Physics, TU Wien, 1040 Vienna, Austria}%

\author{Neven Bari\v{s}i\'{c}}
\affiliation{%
    Institute of Solid State Physics, TU Wien, 1040 Vienna, Austria}%
\affiliation{%
    Department of Physics, Faculty of Science, University of Zagreb, Bijeni\v{c}ka 32, HR-10000 Zagreb, Croatia}%

\author{Karsten Held \orcidlink{0000-0001-5984-8549}}
\affiliation{%
    Institute of Solid State Physics, TU Wien, 1040 Vienna, Austria}%

\date{\today}
    
\begin{abstract}
  We find a strongly  enhanced entanglement within the pseudogap regime of the Hubbard model. This entanglement
  is estimated from  the  quantum Fisher information and, avoiding the ill-conditioned analytical continuation, the quantum variance.  Both are lower bounds for the actual entanglement and can be calculated  from the
  (antiferromagnetic) susceptibility, obtained here with the
    dynamical vertex approximation.
    Our results \kh{qualitatively} agree with experimental neutron scattering experiments for various cuprates.
    Theory  predicts a $\ln(1/T)$ divergence of the entanglement for low temperatures $T$, which is however cut-off by the onset of superconductivity.
%        (D\textGamma A), a non-local extension to the dynamical mean-field theory. The pseudogap, a partially gaped electronic state, is observed near the superconducting transition in cuprates and nickelates. Leveraging D\textGamma A, we compute the quantum variance—a lower bound to the quantum Fisher information—from the spin susceptibility directly on the imaginary Matsubara axis. By circumventing the need for ill-controlled analytical continuation, our approach provides a robust framework for probing entanglement depth. The results show good agreement with experimental data. Additionally, Ornstein-Zernike fits provide analytical insights.}
\end{abstract}
\maketitle

%%%%%%%%%%%%%%%%%%%%%%%%%%%%%%%%%%%%%%%%%%%%%
%\begin{figure}[t]
%    \centering
%    \includegraphics[width=0.95\linewidth]{Plots/graphic_aper_draft.pdf}
%    \caption{The pseudogap is highly entangled. In an idealized world with pure states, measurement of one single spin degree of freedom reveals information about at least 4 more spins. }
%    \label{fig:Explanation}
%\end{figure}

\textsl{Introduction --}
%\section{Introduction}
%Entanglement is one of the key manifestations of quantum mechanics, as such, entanglement in the Fermi-Hubbard model has received much attention in recently \cite{roosz_two-site_2024,bera_dynamical_2024,gu_entanglement_2004,vafek_entanglement_2017,ehlers_entanglement_2015,bellomia_quantum_2024,abaach_long_2023,lo_schiavo_quantum_2023,demidio_universal_2024,walsh_local_2019,walsh_entanglement_2020,gauvin-ndiaye_mott_2023,wang_entanglement_2023,bellomia_quasilocal_2024}.
Measuring entanglement in macroscopic solid-state systems 
is extremely challenging as many particles participate
and are measured simultaneously.
Only very recently, first  entanglement witnesses
such as the  one- and two-tangle \cite{scheie_witnessing_2021,laurell_quantifying_2021} and the  quantum Fisher information (QFI) \cite{Mathew2020,laurell_quantifying_2021,Scheie2024,mazza_quantum_2024}
have been determined  in antiferromagnetic spin chains~\cite{Mathew2020,scheie_witnessing_2021,laurell_quantifying_2021},
%[Cu($\mu$-C$_2$O$_4$)(4-aminopyridine)$_2$(H$_2$O)]$_n$  \cite{Mathew2020},
%KCuF$_3$  \cite{scheie_witnessing_2021}
%and CS$_2$CoCl$_4$ \cite{laurell_quantifying_2021},
triangular-lattice  quantum spin liquids
%KYbSe$_2$
\cite{Scheie2024}, 
heavy fermion systems
%Ce$_3$Pd$_{20}$Si$_6$
\cite{mazza_quantum_2024}\kh{, and --more closely related to our work-- cuprates in the strange metal regime \cite{balut2025quantumfisherinformationreveals}.}
For a  review, see \cite{laurell_witnessing_2024}.
In this context, the QFI is particularly handy because it connects entanglement depth \cite{pezze_entanglement_2009,hyllus_fisher_2012} to the susceptibility \cite{hauke_measuring_2016}, making it  accessible in neutron scattering experiments. \rep{For  general reviews and further  (typically experimentally not accessible)   entanglement measures in many-body systems, we refer the reader to \cite{Amico_2008,Frerot_2023}.}

The QFI is given by the following integral over the imaginary part of the susceptibility $\chi(\omega)$ \cite{hauke_measuring_2016}:
\begin{equation}
  F_{\textrm Q}=\frac{4}{\pi}\int_{0}^{\infty}d\omega\tanh\left(\frac{\omega}{2T}\right)\textrm{Im}\left[\chi\left(\omega\right)\right]
  \label{Eq:QFI}
\end{equation}
with a temperature ($T$) cut-off for the lowest frequencies $\omega$.
If $F_{\textrm Q}>m$, the system is at least $(m\!+\!1)$-partite entangled \cite{hauke_measuring_2016}.
That means, if measuring one particle (or say its spin), $m$ other particles (degrees of freedom) are modified through the collapse of the wave function.
%\khc{maybe D}
For neutron scattering experiments, it  is challenging to properly subtract the background and to include
the large-$\omega$ tail of the magnetic susceptibility when evaluating \cref{Eq:QFI}. On the theory side,
the reliance on real frequencies poses a significant obstacle for many
numerical many-body methods where it necessitates  an ill-conditioned analytical continuation.

The temperature cut-off of \cref{Eq:QFI} 
reflects the quantum nature of entanglement. The
``classical'' $\omega= 0$ susceptibility that %\rep{\sout{diverges at a finite temperature phase transition}}
signals ordering  but not entanglement%\rep{\sout{, and}}  
must not be included in $F_Q$ \rep{\cite{frerot_reconstructing_2019}}. 
\rep{This  $\omega= 0$ component even diverges at a finite-temperature phase transition  irrespectively whether 
we have a classical or quantum model, and thus
does not necessarily reflect any  entanglement. Quantum correlations can remain short ranged while the correlation length and the $\omega= 0$ susceptibility  diverge \cite{Grover_2020}.}
This is the reason why
quantum critical systems and quantum spin liquids were hitherto at the focus of QFI research \cite{hauke_measuring_2016,Mathew2020,scheie_witnessing_2021,laurell_quantifying_2021,Scheie2024,mazza_quantum_2024}.

As for theory, entanglement in many-body solid state systems has traditionally focused on spin systems. Only recently also
strongly correlated electron systems such as the Hubbard model received more attention
\cite{roosz_two-site_2024,bippus2025_2D,Rohshap_2025_QTTD,bera_dynamical_2024,gu_entanglement_2004,vafek_entanglement_2017,ehlers_entanglement_2015,bellomia_quantum_2024,abaach_long_2023,lo_schiavo_quantum_2023,demidio_universal_2024,walsh_local_2019,walsh_entanglement_2020,gauvin-ndiaye_mott_2023,wang_entanglement_2023,bellomia_quasilocal_2024}. Different numerical methods have been employed for studying different entanglement measures and witnesses, but research is still in its infancy.  A more accurate modelling of cuprate superconductors
requires more than the single  band of the Hubbard model, which is quite obvious from the fact  that
holes generated by chemical doping go into the oxygen orbitals \cite{Tranquada1987,barisic_high-t_ccuprates_2022} with the copper orbitals remaining close to half filling. Nonetheless, the Hubbard model is the most commonly studied model for cuprates \cite{Gull2015}, as an effective model
and because \kh{it describes} a single Fermi surface of mixed oxygen and copper character \kh{as observed in experiment}.

\begin{figure}[tb!]
    \centering
    \includegraphics[scale=0.99]{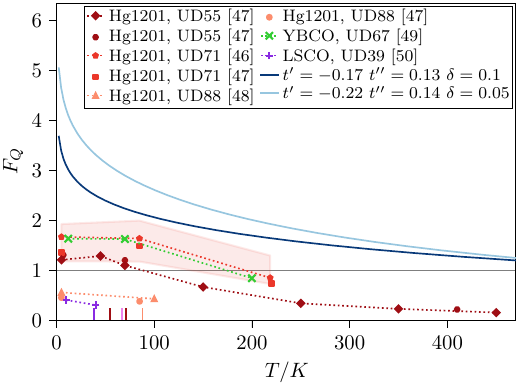}
    \caption{ Entanglement as witnessed by the QFI, comparing  the Hubbard model at two different dopings $\delta$ and
      hopping parameters (solid lines; $t={-0.45}\,$eV and $U=8t$ for both) to that calculated from  various neutron scattering experiments
      \cite{footnoteexp}.
      The experimental data includes large uncertainties, exemplarily indicated by the red shaded range for
        Hg1201 UD71. 
        \kh{Importantly, the experimental QFI only includes the \fb{in}elastic contribution, not the quasielastic contribution which is relevant in theory even with the QFI constraint $\omega>T$.} 
        %\mk{[MK: inelastic? While I am not sure which data is actually used to evaluate QFI, for example, the author of Ref. [38] mentioned they performed an inelastic neutron-scattering study.]}
        %\khc{yes, was a typo}
        %      we display an approximation of the error propagation for Hg1201 UD71 with the shaded region.
        Small vertical lines indicate superconducting critical temperatures ($T_c$) of each material\fb{; and the gray \ak{horizontal} line indicates the onset of bipartite entanglement.}  }
    \label{fig:Experimental}
\end{figure}

In this letter, we report that the pseudogap (PG)  regime of the two-dimensional
Hubbard model shows a \fb{significant} entanglement as measured by the QFI and the  quantum variance (QV) \cite{frerot_quantum_2016, frerot_reconstructing_2019} which we calculate  directly from imaginary (Matsubara) frequencies  $i \omega_n$ as {[see Eq.~(S3) of the Supplemental information \cite{SM}]}
\begin{equation}
  I_{\textrm{Q}}=
  %\frac
  {8T}
  %{\beta}
  \sum_{n=1}^{\infty}\chi(i\omega_n) \leq   F_{\textrm{Q}}  \; .
  \label{Eq:IV}
\end{equation}
The QV $I_{\textrm Q}$ is  a lower bound to the QFI $F_{\textrm{Q}}$ so that
{$I_{Q}>m$} also implies $F_{\textrm{Q}}>m$ and  at least $(m+1)$-partite entanglement.  Applying an Ornstein-Zernike fit, we also obtain the QFI and show that the entanglement of the pseudogap regime grows as $\ln (1/T)$.
Our results for the Hubbard model compare favorably to neutron scattering
experiments for various cuprate superconductors within the pseudogap regime, see Fig.~\ref{fig:Experimental}, the main result of our paper. One reason for the larger theoretical QFI is that the $\ln (1/T)$ divergence is cut-off by the onset of superconductivity. Further\rep{: (i)} The  contribution from the quasielastic peak, important in theory, is missing in neutron scattering experiments due to the opening of a spingap around $T_c$. \rep{(ii) To remove the background, the spectrum at a larger temperature is subtracted. These deviations appear to be more relevant than the uncertainties of the experimental error bars, which we highlight for Hg1201 UD71 as red shaded area. Moreover, the cut-off of the frequency integral leads to an underestimation of the value. We expect the deviation to be small, based on the theoretical observation of the summation error \cite{SM}.}  In any case, theory and experiment together demonstrate an enhanced entanglement within the pseudogap regime of cuprate superconductors.

%%%%%%%%%%%%%%%%%%%%%%%%%%%%%%%%%%%%%%%%%%%%%
\textsl{Model and Method --} 
%\section{Model and Method}
For investigating entanglement in the PG regime of cuprates theoretically, we consider the two-dimensional Hubbard model as a minimal model. Its Hamiltonian reads
\begin{equation}
  H = \sum_{ij,\sigma} t_{ij}  
  c^{\dagger}_{i,\sigma}c_{j,\sigma}^{\phantom{\dagger}}
  + U\sum_{i} n_{i,\uparrow}n_{i,\downarrow}\; .
\end{equation}
Here,  $c_{i,\sigma}^{\dagger}$ and  $c_{i,\sigma}^{\phantom{\dagger}}$  are the creation and annihilation operators for an electron on lattice site $i$ with spin $\sigma$;
 $n_{i,\sigma} = c_{i,\sigma}^{\dagger}c_{i,\sigma}^{\phantom{\dagger}}$ is the occupation number operator. 
We limit the hopping $t_{ij}$  to nearest, next-nearest and next-next-nearest neighbor hoppings $t$, $t'$, and $t''$, respectively;
 $U$ is the onsite Coulomb repulsion.
We use two sets of hopping parameters, $t'=-0.17t$, $t''=0.13t$ and
$t'=-0.22t$, $t''=0.14t$, that cover a typical spread among different cuprates including  HgBa$_2$CuO$_4$ (Hg1201) \cite{Nichiguchi2013,Nichiguchi2013},
but keep  {$t=-{0.45}\,$}eV fixed as it merely sets the overall energy scale. 
We vary the  hole doping $\delta$ \kh{(measured relative to half-filling)} and $U$.
%\kh{Shallw we in the spirit of the cuprates have $\delta$ instead of n, and also plot Fig. 2a invertexd starting like the cuprate phase diagrams?}

To solve the Hubbard model, we  use dynamical mean-field theory (DMFT) \cite{Georges1996} as implemented in {\sc w2dynamics} \cite{Wallerberger_CompPhysComm_2019_w2dynamics} as a first step   and subsequently include non-local correlations beyond DMFT  by 
the  dynamical vertex approximation (D\textGamma A) in its ladder variant with $\lambda$ correction \cite{Toschi2007,rohringer_diagrammatic_2018,Kitatani2022}. This
faithfully reduces the overestimated DMFT susceptibilities \cite{rohringer_diagrammatic_2018} to values that well agree with other numerical approaches \cite{Schaefer2021}. 
\rep{Within D\textGamma A, the local but dynamical DMFT vertex is used to build  non-local generalized susceptibilites through the Bethe-Salpeter equation. For the Hubbard model,  momentum-dependent and frequency-dependent spin fluctuations are here dominating.  Subsequently,
 via the Schwinger-Dyson equation,   a momentum-dependent self-energy is constructed 
 from these generalized susceptibilities or the corresponding  full vertex \cite{rohringer_diagrammatic_2018}. Thereby, in the two-dimensional Hubbard model
 spin fluctuations lead to the opening of the PG \cite{Katanin2009,rohringer_diagrammatic_2018}.}

For the dominant antiferromagnetic (AFM) wave-vector ${\mathbf Q}=(\pi\pm\Delta,\ \pi)$ or $(\pi,\pi\pm\Delta)$,
the D\textGamma A magnetic susceptibility $\chi(i\omega_n)$ at bosonic Matsubara frequencies $\omega_n=2\pi T n$ 
is used to compute the QV directly  using \cref{Eq:IV}. \rep{It has been shown, that the susceptibility at a specific wave-vector can be used to witness entanglement \cite{wang_2025}. For our choice of the local operator, $O=\sum_j e^{i\mathbf{q}\mathbf{r}_j}S_j$, maximal entanglement is witnessed at $\mathbf{q}=\mathbf{Q}$.}
In the Supplemental information \cite{SM}, we show how the incommensurability $\Delta$ depends on the doping $\delta$. {In the main part we only show $\Delta=0$ results (except of the LSCO experiment).
Note that \cref{Eq:IV} remains a proper lower bound for entanglement, even if the summation is only performed up to a finite $n$; and we apply a cut-off at half the bandwidth,
$\omega_n< {W}/{2} = 4( t+t'')$ if not stated otherwise.
For calculating also the QFI, we fit
%note that the QV is dominated by the low frequency behavior of
$\chi(i\omega_n)$  to the Ornstein Zernike (OZ) form
\begin{equation}
  \chi(i\omega_n) = \frac{A}{\xi^{-2}+\gamma\left|i\omega_n\right|}  \;
  \label{Eq:OZ}
\end{equation}
and integrate \cref{Eq:QFI} with  the analytical continuation thereof and cut-off $W/2$. The temperature dependence 
of the amplitude $A$, correlation length $\xi$ and (Boson)
damping factor $\gamma$
as fitted to the D$\Gamma$A is shown in the Supplemental Material \cite{SM}.
%we will provide an analytical approximation to the QFI.
%As already mentioned, if the QFI or the QV is larger than $m\in\mathbb{N}$ this indicates that the system is at least $(m+1)$-partite entangled.

For the experimental data we digitized the magnetic susceptibility   from the literature \cite{chan_commensurate_2016,tang_neutron_2018,chan_hourglass_2016,fong_spin_2000,christensen_dispersive_2004}
} with the background already subtracted and frequency ranges starting from between $0$ to $40$ meV and ranging up to $20-190$ meV. A simple integration with the trapezoidal rule %\kh{\sout{(also known as the trapezoid rule or trapezium rule)}} 
has been employed. In the same way as for the theory results, we
calculate the QFI for the leading AFM wave vector ${\mathbf Q}$.
\kh{To this end,} \fb{experimental susceptibilitites are rescaled by the g-factor $g=2$, $\kh{\chi^{zz}=}\frac{1}{g^2}\chi^{\textrm{Exp}}$, to ensure the correct normalization of the QFI \cite{SCHEIE2025}.}
\kh{Note that the subtraction of the background is very complicated in most experiments for both low and \fb{high} frequencies,  possibly leading to an underestimation of the QFI.
}

\kh{Even more importantly, we did not find properly normalized neutron data for the low frequency quasielastic peak which means that a relevant contribution to the QFI is missing. This certainly is a factor why the experimental QFI is smaller than the D$\Gamma$A calculated one. The latter to a large part indeed  stems from the quasielastic peak, see Eq.~(\ref{Eq:OZ}).}

%%%%%%%%%%%%%%%%%%%%%%%%%%%%%%%%%%%%%%%%%%%%%

\begin{figure*}[tb]
    \centering
    \includegraphics[scale=1.0]{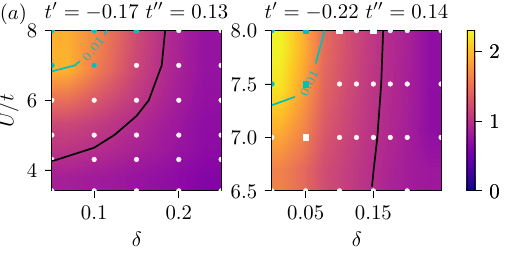} %\hspace{-.5cm}
    \includegraphics[scale=1.0]{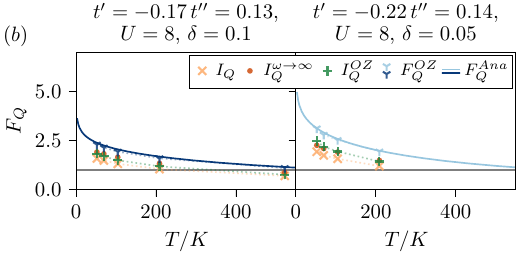}%trim={0 0 .9cm 0}
    \caption{(a) QV as a function of Coulomb repulsion $U$ and doping $\delta$  for two sets of $t',\ t''$ and $T = 52\,$K. \fb{Black lines indicate \ak{the region} where \ak {at least a bi-partite} entanglement is detected,} \kh{(cyan/white) dots and squares represent computed data points (inside/outside)}  the PG regime; for the squares, the spectral function is shown in Fig~\ref{fig:PG_opening}; and the cyan line displays a \ak{$\chi^{-1} (i\omega =0)=0.01$} contour that serves as a proxy for the onset of the PG \cite{kitatani_optimizing_2023}. 
    (b) Comparison of  QV from (i) sum over Matsubara frequencies ($I_{\textrm Q}$), (ii) extrapolated to infinite frequencies  ($I_{\textrm Q}^{\omega\rightarrow \infty}$), (iii) from OZ fit  ($I_{\textrm Q}^{\textrm OZ}$), and the QFI
    from (iv) the  OZ fit  ($F_{\textrm Q}^{\textrm OZ}$) and its analytical evaluation  ($F_{\textrm Q}^{\textrm Ana}$).
    In the left figure, we display $t'=-0.17\kh{t},\ t''=0.13\kh{t}$ at $U=8t$ and $\delta=0.1$ while in the right panel $t'=-0.22\kh{t},\ t''=0.14\kh{t}$ at $U=8t$ and $\delta=0.05$ is shown. Dotted lines are guides for the eye \fb{and the \ak{horizontal} gray line indicates the barrier for bipartite entanglement.} %\mk{[MK:the line color of the analytical formula for $t'=0.17,t''=0.13$ looks different to the legend...]}\khc{can we have light and dark blue lines simultaneously in the legend?}\fb{I have added the dark blue symbols to the legend, is this fine this way? }
    }
    \label{fig:Phase_Diagram}
\end{figure*}
%\end{document}
\begin{figure*}[tb]
    \centering
    \includegraphics[width=\linewidth]{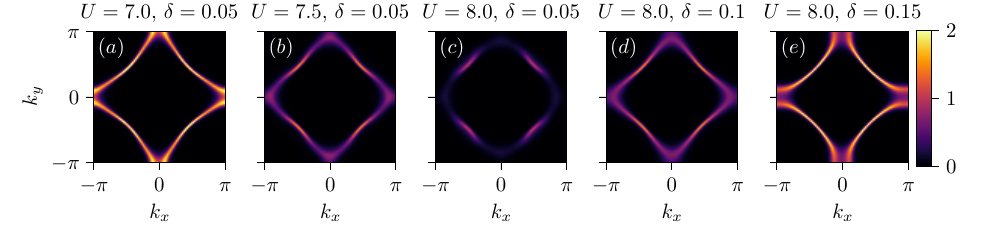} %\hspace{-.5cm}
    \caption{Momentum-resolved spectral function around the Fermi energy $-\frac{\beta}{\pi}G\left(\mathbf{k},\tau=\beta/2\right)$ showing the pseudogap opening with increasing $U$ from (a) to (c) at fixed $\delta=0.05$ and closing again at fixed $U=8.0$ with increasing hole doping $\delta$ from (c) to (e). The panels correspond to the squares in Fig.~\ref{fig:Phase_Diagram} (a) with $t'=-0.22\kh{t},\ t''=0.14\kh{t}$, $U=8t$ and $T=5\kh{2}$K.}
    \label{fig:PG_opening}
\end{figure*}

\textsl{Results --}
%\section{Results}
In Fig.~\ref{fig:Phase_Diagram}~(a) we show  the QV as a function of 
$U$ and $\delta$  for the two sets of hopping parameters $t'$, $t''$ considered. The QV drastically increases towards large $U$ and small $\delta$, within the  PG regime.  The cyan dots indicate those points for which we observe a  PG in the spectral function, see Supplemental Material \cite{SM}. Here, within the PG regime the QV becomes \fb{ at least three-partite for the studied temperature $T=52\,$K.}
\rep{Hence, in the PG regime, spins can not be considered to be independent, and measurements performed on a single spin will strongly influence multiple other spins.}

In the PG region, also the static susceptibility $\chi(\omega=0)$ that does not contribute to the entanglement is large ({cyan} iso-value line). Indeed the largeness of  $\chi(\omega=0)$ has been taken as a proxy for the PG \cite{kitatani_optimizing_2023}\rep{, and empirically agrees very well with the opening of the PG \cite{SM}}. Whether we have a pseudogap or not is shown exemplarily in Fig.~\ref{fig:PG_opening} and 
Supplemental Material \cite{SM}. Here,   
\begin{equation}
-\frac{1}{\pi T}G\left(\mathbf{k},\tau=1/(2T)\right)=\frac{1}{2\pi T}\int\text{d}\omega\frac{A\left(\mathbf{k},\omega\right)}{\cosh\left(\omega/(2T)\right)}
\end{equation}
is an average of the spectral function $A(\mathbf{k},\omega)$  over a frequency range $\pm T$ around the Fermi energy, that avoids the ill-conditioned analytical continuation.

Fig.~\ref{fig:Phase_Diagram}~(b) presents
the temperature dependence of the QV at fixed $U$ and $\delta$. In this figure, we also plot  the QV summed over all  Matsubara frequencies $I_{\mathrm Q}^{\omega\rightarrow \infty}$. We see that this full integration actually only leads to a minor enhancement of $I_{\mathrm Q}$ compared to the summation up to the half-bandwidth cut-off. Further, we fit $\chi(\omega_n)$ to the OZ form, and integrate  $I_{\mathrm Q}^{\textrm{OZ}}$ on the real frequency axis with the OZ fit according to Eq.~(S7) of the Supplemental Material \cite{SM}.  The good agreement between  $I_{\mathrm Q}^{\omega\rightarrow \infty}$ and
$I_{\mathrm Q}^{\textrm{OZ}}$ validates the quality of the OZ fit \cite{footnoteQ}.
Independent of the applied method, we observe that bi-partite entanglement arises around room temperature or above, \rep{which is also the temperature range at which the PG first opens in cuprates \cite{Louis_2018}.}

Let us now turn to the QFI $F_{\mathrm Q}^{\mathrm OZ}$ which is integrated on the real frequency axis with the OZ fit as well.  As mathematically required,
the QFI is higher than the QV, however only by about \kh{$1/2$}.
Given the fact that the QFI by itself is only a rough lower bound of the actual entanglement this additional underestimation of the entanglement by  \kh{$1/2$} appears bearable, in particular as the QV and QFI show the same overall trends.

Finally, we arrive in the Supplemental Material \cite{SM} at an  analytic form of the QFI which holds in the limit of low $T$ (and large correlation length $\xi$ that is approached for low $T$): 
%\fb{The approximation only works for large $\xi$, I would rather write: which holds in the limit of large correlation lengths $\xi$ as approached at low temperatures $T$}:
\begin{equation}
	F_Q \approx \frac{4}{\pi }\frac{A}{\gamma}  \left[ 1 + \ln\left(\frac{W\fb{/2}}{2T}\right)\right]\;.
\end{equation}
This analytical form is plotted as a {solid} line in Fig.~\ref{fig:Phase_Diagram}~(b) and  is in good agreement with the QFI from the OZ fit in the whole $T$-range considered. 
It shows a $\ln(1/T)$ divergence for low temperatures.  This divergence directly stems from the leading $1/\omega$ behavior for small $\omega$  in the OZ form \cref{Eq:OZ}, with temperature $T$ acting as a cut-off. (There is another cut-off given by the
inverse correlation length $\xi^{-1}$ which is however smaller in two dimensions where  $\xi$ grows exponentially in $1/T$.)
Note that the prefactor $A/\gamma$ approaches a constant for $T\rightarrow 0$, see  Supplemental Material \cite{SM}.
Physically, the  term $\gamma \omega$ in \cref{Eq:OZ} describes the overdamping of the bosonic
susceptibility mode which is characteristic  of strongly correlated electron systems.
It describes the dynamical (quantum part of the) susceptibility in the frequency regime that is most relevant
for the QFI estimate of entanglement.
\rep{In contrast to the vicinity of quantum critical points, the QFI as a lower bound for entanglement is --to a first approximation-- independent of the correlation length $\xi$ within the PG regime. At a quantum critical point, the diverging correlation length would also lead to a diverging QFI with the same critical scaling \cite{hauke_measuring_2016}. }

In \cref{fig:Experimental} we plot this analytical form of the QFI
and compare it to  neutron scattering susceptibility data extracted from various experiments for cuprate superconductors in the PG regime \cite{chan_commensurate_2016,tang_neutron_2018,chan_hourglass_2016,fong_spin_2000,christensen_dispersive_2004}. Both actually agree very well for  HgBa$_2$CuO$_{4+x}$  (Hg1201)  at an under-doping corresponding to $T_c=55$\,K  (UD55) and UD77\kh{,  and to}  YBa$_2$Cu$_3$O$_{6+x}$ (YBCO) UD67, demonstrating that in cuprates and the Hubbard model alike we have at least a \fb{bi-partite entanglement as $F_Q>1$ and, as discussed, experimental values are expected to underestimate the actual value.} At least one reason why 
La$_{2-x}$Sr$_x$CuO$_4$ (LSCO) UD39 and Hg1201 UD88 have a lower QFI is that these are closer to optimal doping where the pseudogap gets weaker, and also in theory the QFI is considerably smaller.

An important difference to theory is that the experimental QFI saturates for low $T$. This saturation coincides with entering the superconducting phase (small vertical lines in Fig.~\ref{fig:Experimental}). An aspect that is not captured by our theoretical calculation which does not allow for symmetry breaking.

\kh{For the bilayer cuprate YBCO, we show in Fig.~\ref{fig:Experimental} the QFI for ${\mathbf Q}=({\pi,\pi,0})$ as we consider in our calculation only a single layer.
\rep{We emphasize, that in this 3D material at} ${\mathbf Q}=({\pi,\pi,\pi})$ the QFI is twice as large, i.e., $F_Q\gtrsim 5$, which shows that there is not only entanglement within the layers but also between them \cite{fong_spin_2000}.}

Finally, we note that the Quantum Fisher Information (QFI) can also be computed from other normalizable response functions, such as density fluctuations. For the Hubbard model and parameters at hand, the QFI from the density–density susceptibility remains smaller than one, meaning that no entanglement is observed.

%%%%%%%%%%%%%%%%%%%%%%%%%%%%%%%%%%%%%%%%%%%%%
\textsl{Discussion and Conclusion--}
%\section{Discussion and Conclusion}
%\khc{Maybe Fig. for the discussion?}
The Mermin-Wagner theorem \cite{Mermin66} prevents AF order for any finite temperature on an infinite and  strictly two dimensional lattice.
There is however a $T=0$ quantum critical line of AFM order up to a critical doping $\delta_c$.
In actual cuprates, finite sample  sizes and even a weak hopping perpendicular to the CuO$_2$ planes
lead to AFM order of the parent compounds \cite{Palle_2021}. Nonetheless, AFM is strongly suppressed and restricted to much smaller dopings than in a three dimensional system.
This leads to largely enhanced  AFM fluctuations 
above the hypothetical quantum critical line of the ideal
two-dimensional system.
These enhanced AFM fluctuations are often considered to be the microscopic origin for the  pseudogap and unconventional superconductivity in cuprates \cite{Scalapino12}.

Here, we have shown that these strong antiferromagnetic fluctuations, with the $\omega=0$ static contribution explicitly taken out, also mean %\kh{an  \sout{strongly}} 
enhanced entanglement within the pseudogap regime of the Hubbard model and in cuprate superconductors. This entanglement grows like $\ln (1/T)$ and  only saturates when entering the superconducting phase. \rep{The multi-partite entanglement was found for the 
antiferromagnetic  spin susceptibility. We thus expect that when measuring one spin
in the PG phase of cuprates, other spins (most likely neighboring ones) will be strongly affected through the collapse of the quantum mechanical wave function or density matrix.}

The entanglement \rep{in the PG phase of the Hubbard model and in cuprates} as deduced from the  QFI and the QV, which is more generally accessible in theory, is comparable to that of other materials with strong entanglement such as quantum spin liquids and heavy fermion systems. \kh{As for the still lower experimental QFI in  cuprates compared to theory, the ball is now in the court of neutron scattering experiment, which hitherto  had a focus on the elastic peak. However,
 the quasieleastic peak at $\omega>T$ is very important for the QFI.}

%cuprates are strongly

%how to faithfully measure multipartite entanglement from numerical methods working on the imaginary Matsubara frequency axis. Applying this, we were able to show strongly enhanced entanglement in the pseudo gap regime of cuprates, a behavior which has thus far only been observed close to quantum critical points. Based on the Ornstein Zernike form for susceptibilities we have provided an analytical equation for the entanglement in the pseudo gap. Our results are in good agreement with experimental data. 
%Future research should be focused on two main aspects. Having in mind the limitations of the experimental data available, we urge for the experimental measurement of susceptibilities at a higher frequency resolution to allow for more accurate results.
%On the theoretical side, more work will be necessary to understand the structure of the entanglement in the PG. However, the QFI and QV as measures of multi-partite entanglement are not suited for this task and density-matrix based measures will be necessary \cite{roosz_two-site_2024}.

%%%%%%%%%%%%%%%%%%%%%%%%%%%%%%%%%%%%%%%%%%%%%
\begin{acknowledgments}
\textsl{Acknowledgments --}
We thank Fakher Assaad, Frederico Mazza, \rep{Dongwook Kim,} and Gerg\"o Roosz for very helpful discussions. Support by the Spezialforschungsbereich (SFB) Q-M\&S of the Austrian Science Funds (FWF; project DOI 10.55776/F86) is gratefully acknowledged. \mk{M. K. appreciates the support from the Grant-in-Aid for Scientific Research (JSPS KAKENHI) Grant No. JP24K17014.} A.K. acknowledges support by the FWF project V~1018 (grant DOI 10.55776/V1018)  Calculations have been done in part on the Vienna Scientific Cluster (VSC). For the purpose of open access, the authors have applied a CC BY-NC-SA public copyright license to any Author Accepted Manuscript version arising from this submission.
\end{acknowledgments}

\section*{DATA AVAILABILITY}
The data that support the findings of this article are openly available \cite{data}.

\bibliography{references.bib}

%\import{./}{Supplemental.tex}

\end{document}

\title{Supplemental Material: Entanglement in the pseudogap regime of
  %the Hubbard model and
  cuprate superconductors}

\author{Frederic Bippus}
\affiliation{%
    Institute of Solid State Physics, TU Wien, 1040 Vienna, Austria}%

\author{Juraj Krsnik}
\affiliation{%
    Institute of Solid State Physics, TU Wien, 1040 Vienna, Austria}%

\author{Motoharu Kitatani}
\affiliation{%
    Department of Material Science, University of Hyogo, Ako, Hyogo 678-1297, Japan}%
\affiliation{%
    RIKEN Center for Emergent Matter Sciences (CEMS), Wako, Saitama 351-0198, Japan}%

\author{Luka Ak\v{s}amovi\'{c}}
\affiliation{%
    Institute of Solid State Physics, TU Wien, 1040 Vienna, Austria}%
    
\author{Anna Kauch}
\affiliation{%
    Institute of Solid State Physics, TU Wien, 1040 Vienna, Austria}%

\author{Neven Bari\v{s}i\'{c}}
\affiliation{%
    Institute of Solid State Physics, TU Wien, 1040 Vienna, Austria}%
\affiliation{%
    Department of Physics, Faculty of Science, University of Zagreb, Bijeni\v{c}ka 32, HR-10000 Zagreb, Croatia}%

\author{Karsten Held}
\affiliation{%
    Institute of Solid State Physics, TU Wien, 1040 Vienna, Austria}%

\date{\today}
    
\begin{abstract}{
    Here, we recapitulate and define $m$-partite entanglement in   Section~\ref{SSec:entanglement};
    rewrite the quantum variance on the Matsubara axis in
Section~\ref{SSec:QV}; discuss the Ornstein Zernicke fit in
Section~\ref{SSec:OZ}; derive an analytical expression for the quantum Fisher information in
Section~\ref{SSec:ana}; discuss the crossover from commensurability to incommensurability in
Section~\ref{SSec:incom}; and demonstrate opening of the pseudogap in the spectral function in
Section~\ref{SSec:PG}.}

\end{abstract}
\maketitle

%%%%%%%%%%%%%%%%%%%%%%%%%%%%%%%%%%%%%%%%%%%%%
%begin{comment}
\section{Multipartite Entanglement}
\label{SSec:entanglement}
For the sake of completeness, let us start by recalling the definition of $m$-partite entanglement and $m$-producibility.
A quantum state 
\begin{equation}
    \left| \psi_i \right\rangle = \left| \phi_1 \right\rangle \otimes \left| \phi_2 \right\rangle \otimes ... \otimes \left| \phi_l \right\rangle
\end{equation}
is $m$-producible if each $\left| \phi_i \right\rangle$ is a state of at most $m$ spins or other qbit  degrees of freedom. 
The mixed state $\rho = \sum_i p_i \left| \psi_i \right\rangle\left\langle \psi_i \right|$ is $m$-producible if all $\left| \psi_i \right\rangle$ are at most $m$-producible.
Vice versa, if none of the states $\left| \psi_i \right\rangle$ is $m$-producible, the system is at least $(m+1)$-partite entangled \cite{guhne_multipartite_2005, hyllus_fisher_2012, pezze_entanglement_2009}. At least one 
   $\left| \phi_i \right\rangle$ must involve $m+1$ qbits.

In general, this criterion does not distinguish if there is just a single highly entangled state in the system or if multiple states with the same entanglement exist \cite{szalay_alternatives_2024}. However, translational invariance in our system requires that each spin or charge degree of freedom
  is included in at least one state with high entanglement.

  Please further note, that there are four states (empty, doubly occupied, spin-$\uparrow$ and -$\downarrow$) for each site of the Hubbard model, i.e., two qbits or  a spin and charge degree of freedom.
  Since we measure entanglement by the QFI calculated from the spin susceptibility, it is however
  clear that the entanglement we observe mainly originates from the spin degrees of freedom.
%end{comment}

%%%%%%%%%%%%%%%%%%%%%%%%%%%%%%%%%%%%%%%%%%%%%
\section{Quantum Variance as a Lower Bound to the Quantum Fisher Information}

\label{SSec:QV}
As already mentioned in the main text, the quantum Fisher information (QFI) is related to the dynamic susceptibility $\chi(\omega)$ as follows \cite{hauke_measuring_2016}:
\begin{equation}
F_{Q}=\frac{4}{\pi}\int_{0}^{\infty}d\omega\tanh\left(\frac{\omega\beta}{2}\right)\textrm{Im}\left[\chi\left(\omega\right)\right]
\end{equation}
with inverse temperature $\beta = 1/T$. A value of $F_{Q}\geq m$ indicates that a system is at
least $(m+1)$ partite entangled \cite{hyllus_fisher_2012}. While the QFI is not directly obtainable by numerical methods working with imaginary Matsubara frequencies, we will show that the quantum variance (QV) $I_{\textrm{Q}}$ can be computed directly from Matsubara frequencies.
Like the QFI, the QV  is a lower bound for the multi-partite entanglement \cite{frerot_quantum_2016,frerot_reconstructing_2019,scheie_reconstructing_2024,lambert_classical_2023,laurell_witnessing_2024}.
The relation between QFI and QV becomes apparent by expressing both quantities as sums
which is possible as the $\tanh$ can be expressed by a sum:
\begin{equation}\label{eq_approx_Qf}
\begin{aligned}F_{\textrm{Q}} = & \frac{4}{\pi}\int_{0}^{\infty}d\omega\tanh\left(\frac{\omega\beta}{2}\right)\textrm{Im}\left[\chi\left(\omega\right)\right]\\
= & \frac{4}{\pi}\sum_{n=0}^{\infty}\int_{0}^{\infty}d\omega\frac{4\omega\beta}{\left(2n+1\right)^{2}\pi^{2}+\left(\omega\beta\right)^{2}}\textrm{Im}\left[\chi\left(\omega\right)\right]\\
\geq & \frac{4}{\pi}\sum_{n=1}^{\infty}\int_{0}^{\infty}d\omega\frac{4\omega\beta}{\left(2\pi n\right)^{2}+\left(\omega\beta\right)^{2}}\textrm{Im}\left[\chi\left(\omega\right)\right]\\
\equiv & I_{\textrm{Q}}.
\end{aligned}
\end{equation}
Using $I_Q$, we can further identify   $\omega_{n}=\frac{2\pi n}{\beta}$ as the $n$th Matsubara frequency and
recall  the bosonic Matsubara frequency kernel of the Kramers-Kronigs relation \cite{kaufmann_ana_cont_2021}
\begin{equation}
\textrm{Re}\left[\chi\left(i\omega_{n}\right)\right]=\frac{2}{\pi}\int_{0}^{\infty}d\omega\frac{\omega}{\omega_{n}^{2}+\omega^{2}}\textrm{Im}\left[\chi\left(\omega\right)\right].
\end{equation}
Hence
\begin{equation}\label{Seq_IQ}
\begin{aligned} I_{\textrm{Q}} = & \frac{8}{\beta}\sum_{n=1}^{\infty}\textrm{Re}\left[\chi\left(i\omega_{n}\right)\right]\end{aligned}.
\end{equation}
We have thereby found a lower bound to the quantum Fisher information
$I_{\textrm{QV}} \leq F_{\textrm{Q}}$ expressed directly in bosonic Matsubara
frequencies which enables the detection of genuine $(m+1)$-partite
entanglement if $I_{\textrm{Q}}\geq m$ \cite{methods}. The bound remains genuine also if the sum only goes to a finite $n<\infty$. Like the QFI, the connection between entanglement and QV requires the operators $\hat{O}$ for which the susceptibility $\chi(\omega)=i\int_{0}^{\infty}dt e^{i\omega t}\langle [ \hat{O}(t),\hat{O}]\rangle$ is constructed to be normalized.
From \cref{eq_approx_Qf} one can also compute the QV on the real frequency axis \cite{frerot_reconstructing_2019}. It is advisable to use the sum expression only for low values of $\omega$ and switch to the equivalent form
\begin{equation}
    I_{\textrm{QV}} = \frac{1}{\pi}\int_0^{\beta} d\omega \left(- \frac{2}{\omega\beta} + \coth \frac{\omega\beta}{2}\right)\textrm{Im}\left[\chi\left(\omega\right)\right]
\end{equation}
at large $\omega$. This is due to the slow convergence of the sum  \cref{eq_approx_Qf} at large $\omega$ on the one hand, and the numerical difficulties to compute $-1/x + \coth x$ at low $\omega$ on the other hand. \rep{Regardless of the choice of measure, the cut-off function $-2/\omega\beta + \coth \omega\beta/2$ for the QV and $\tanh \omega\beta/2$ for the QFI vanishes at $\omega=0$, ensuring that the classical contribution at $\omega=0$ is excluded from the integration.}

%%%%%%%%%%%%%%%%%%%%%%%%%%%%%%%%%%%%%%%%%%%%%
\section{Ornstein Zernike Form and Frequency Sums}
\label{SSec:OZ}
\begin{figure}
    \centering
    \includegraphics[scale=1.0]{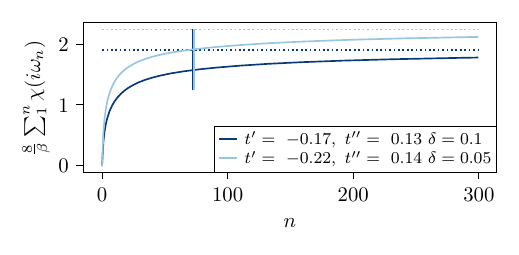}
    \caption{Convergence of the QV  $I_{\textrm Q}=8/\beta \sum_1^n {\rm Re} \chi(i\omega_n)$ for $t'=-0.17,\ t''=0.13$ at $U=8t$ and hole doping $\delta=0.1$ per site and $t'=-0.22,\ t''=0.14$ at $U=8t$. Horizontal line displays the value it converges against, vertical lines indicates the energy scaled summation limit.}
    \label{fig:conv}
\end{figure}

Numerical calculations of the susceptibility are limited to a finite frequency range, nevertheless the QV with such a finite-frequency cut-off  provides a faithful lower bound to multipartite entanglement. In Fig.~\ref{fig:conv} we investigate the impact of this restriction. As one would expect, the main contribution to the QV comes from the first few Matsubara frequencies. Hence, a cutoff at $W/2$ already captures the essence of the QV.

\begin{figure*}
    \centering
    \includegraphics[scale = 1.0]{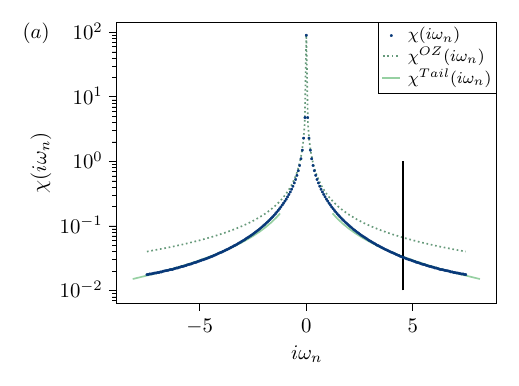}
    \includegraphics[scale = 1.0]{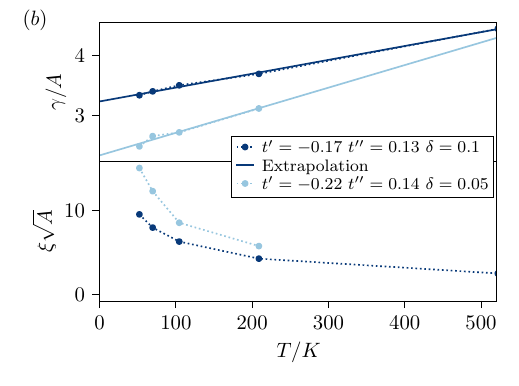}
    \caption{(a) Example of Ornstein Zernike and tail fit for $t'=-0.17,\ t''=0.13$ at $\beta=100$, $U=8t$, and $n=0.90$. A clear overestimation of the tails by the Ornstein Zernike form can be observed. The vertical line indicates the integration limit. 
    (b) Parameters of Ornstein Zernike fit and extrapolation of $\gamma/A$ for $t'=-0.17,\ t''=0.13$ at $U=8t$ and $\delta=0.1$ and $t'=-0.22,\ t''=0.14$ at $U=8t$ and $\delta = 0.05$. Dotted lines are guides for the eye.}
    \label{fig:OZ_Fit}
\end{figure*}

In Fig.~\ref{fig:OZ_Fit}~(a) we show how the susceptibility data has been extrapolated to obtain the results for large frequencies. Here we  assume a decay of $\chi\propto 1/(\gamma \left|i\omega_n\right|+\varepsilon \left|i\omega_n\right|^2)$.
The observation that the first few Matsubara frequencies are dominating the result also justifies the approximation of the QV by using the Ornstein Zernike (OZ) form \cite{hertz_quantum_1976,lohneysen_fermi-liquid_2007,millis_phenomenological_1990}, as a low frequency approximation of the susceptibility:
\begin{equation}\label{eq_oz}
    \chi(i\omega_n) = \frac{A}{\xi^{-2}+\gamma\left|i\omega_n\right|}.
\end{equation}
In Fig.~\ref{fig:OZ_Fit}~(a) we show that this is indeed justified for low frequencies. However, the tails are significantly overestimated as they only fall of as $1/\omega_n$ for the OZ which necessitates the cut-off; otherwise the frequency sums do not converge. While the OZ approximation is not proper for large frequencies, 
Fig.~2~(b) of the main text demonstrates that using the cut-off and the OZ form actually only leads to a minor deviation.
The advantage of the OZ form on the other hand is that it allows us to calculate the QFI and to provide the analytical expression for it that is derived in the next Section.

Further, in  Fig.~\ref{fig:OZ_Fit}~(b) we report the Ornstein Zernicke parameters correlation length $\xi$ and
$\gamma/A$ as obtained from fitting the susceptibility on the imaginary axis to the OZ form.
Here, the susceptibility was calculated with the  dynamical vertex approximation.

%%%%%%%%%%%%%%%%%%%%%%%%%%%%%%%%%%%%%%%%%%%%%
\newpage
\section{Analytical approximation to the Ornstein Zernike Quantum Fisher Information}
\label{SSec:ana}
In this section we derive an approximation to the quantum Fisher information for the Ornstein Zernike susceptibility. By focusing on the strongest contribution with ${\mathbf q}={\mathbf Q}$, we write  its imaginary part as

\begin{equation} \label{eq:OZ_imag}
	\text{Im} \chi^{OZ}(\omega) =  \frac{A\gamma\xi^4\omega}{1 + \gamma^2\xi^4\omega^2 }\;.
\end{equation}
We can now get the QFI by evaluating the integral

\begin{equation} \label{eq:QFI_OZ}
	\int_{0}^{W}d\omega\;\tanh\left( \frac{\omega}{2T}\right) \text{Im}\chi^{OZ}(\omega)\;,
\end{equation}
where we replaced the upper bound of integration with the physically motivated cutoff equal to the bandwidth $W/2$.

Since the integral in Eq.~\eqref{eq:QFI_OZ} can not be obtained in a closed form, we linearize the tanh for low frequencies $(\omega<2T)$ and replace it with 1 for large frequencies $(\omega>2T)$, so that we can approximately write

\begin{equation}
	\begin{split}
			&\int_{0}^{W/2}d\omega\;\tanh\left( \frac{\omega}{2T}\right) \text{Im}\chi(\omega)\\&
			\approx\int_{0}^{2T}d\omega\;\left( \frac{\omega}{2T}\right) \text{Im}\chi(\omega) + \int_{2T}^{W/2}d\omega\;\text{Im}\chi(\omega)=\mathcal{I}_1+\mathcal{I}_2 \;.
		\end{split}
\end{equation}
Here, we also implicitly assume that $2T<W/2$ holds.

We now get

\begin{equation}
	\begin{split}
		\mathcal{I}_1 = \frac{A}{\gamma} - \frac{A}{2T\gamma^2\xi^2} \arctan(\gamma\xi^2 2T )\;,
	\end{split}
\end{equation}
and if the correlation length grows faster than $T^{-1/2}$, $\mathcal{I}_1$ further reduces to $\mathcal{I}_1 \to \frac{A}{\gamma}$ in the low temperature limit.
For the second integral, we can similarly evaluate

\begin{equation}
	\begin{split}
		\mathcal{I}_2 = \frac{A}{2\gamma} \ln \left(\frac{(W/2)^2\gamma^2\xi^4+1}{4T^2\gamma^2\xi^4+1}\right)\;,
	\end{split}
\end{equation}
which again in the $T\to 0$ limit takes a simpler form $\mathcal{I}_2 \to \frac{A}{\gamma}\ln\left(\frac{W/2}{2T}\right)$.

In total, for the  QFI with the OZ form of the susceptibility we approximately get 

\begin{equation}
\begin{split}
    F_Q \approx & \frac{4}{\pi } \left[ \frac{A}{\gamma} - \frac{A}{2T\gamma^2\xi^2} \arctan(\gamma\xi^2 2T ) \right. \\
    &+ \left. \frac{A}{2\gamma} \ln \left(\frac{(W/2)^2\gamma^2\xi^4+1}{4T^2\gamma^2\xi^4+1}\right)\right]\;,
\end{split}
\end{equation}
which is reduced for low $T$ to

\begin{equation}
	F_Q \approx \frac{4}{\pi }\frac{A}{\gamma}  \left[ 1 + \ln\left(\frac{W/2}{2T}\right)\right]\;.
\end{equation}
This approximation is justified as can be seen from
Fig.~2(b) of the main text.
 
%%%%%%%%%%%%%%%%%%%%%%%%%%%%%%%%%%%%%%%%%%%%%
\section{Incommensurate magnetic fluctuations}
\label{SSec:incom}
The results presented in the main text are at low doping within the PG regime where the AFM ordering wave vector ${\bf Q}=(\pi,\pi)$ is still commensurate. However, outside of the PG regime at larger hole doping $\delta$ and lower temperature $T$, the magnetic ordering structure changes to incommensurate ordering ${\mathbf Q}=(\pi\pm\Delta,\pi)$.

\begin{figure*}
    \centering
    \includegraphics[scale=1.0]{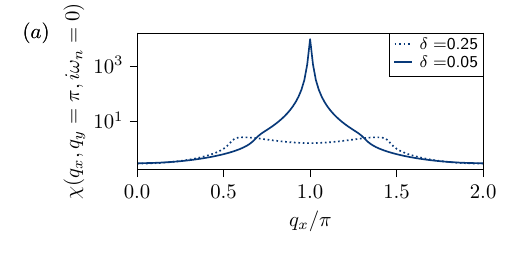}
    \includegraphics[scale=1.0]{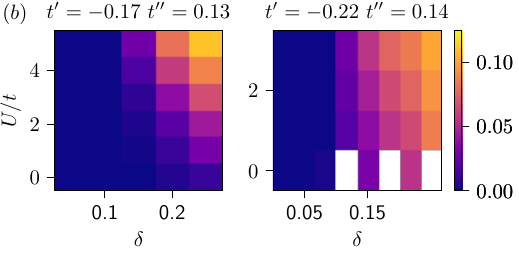}
    \caption{(a)
      Static magnetic susceptibility vs.~momentum $q_x$ in the $x$-direction at fixed $q_y=\pi$, showing the transition
      from dominating commensurate antiferromagnetic fluctuations at $q_x=\pi$ for $\delta=0.05$
      to incommensurate  $q_x\approx \pi\pm0.4\pi$ at $\delta=0.25$.
      The parameters for the Hubbard model are:  $t'=-0.17,\ t''=0.13$, $\beta=100$, and $U=8t$. 
      (b) The resulting difference in QV $I_Q(q=\pi\pm\Delta,\pi) - I_Q(q=\pi,\pi)$ displayed in dependence on $U$ and $\delta$.
    }
    \label{fig:chi_q}
\end{figure*}

Fig.~\ref{fig:chi_q} (a) shows this shift of the maximal susceptibility. For incommensurate ${\mathbf Q}$, the magnetic order is not
following the lattice structure but rather changes over a fraction of the lattice spacing \cite{Dagotto1994}.
Fig.~\ref{fig:chi_q} (b) shows how this incommensurability $\Delta$ as a function of $U$ and doping $\delta$ changes the QV.

A problem is that when ${\mathbf Q}$ changes from commensurate to incommensurate upon lowering $T$ or when
increasing $\delta$
there is a crossover region where  the OZ is not applicable any longer. For the
dopings discussed in the main text, this is however not the case.

%%%%%%%%%%%%%%%%%%%%%%%%%%%%%%%%%%%%%%%%%%%%%
\section{Pseudogap}
\label{SSec:PG}
\begin{figure*}
    \centering
    \includegraphics[scale=0.8]{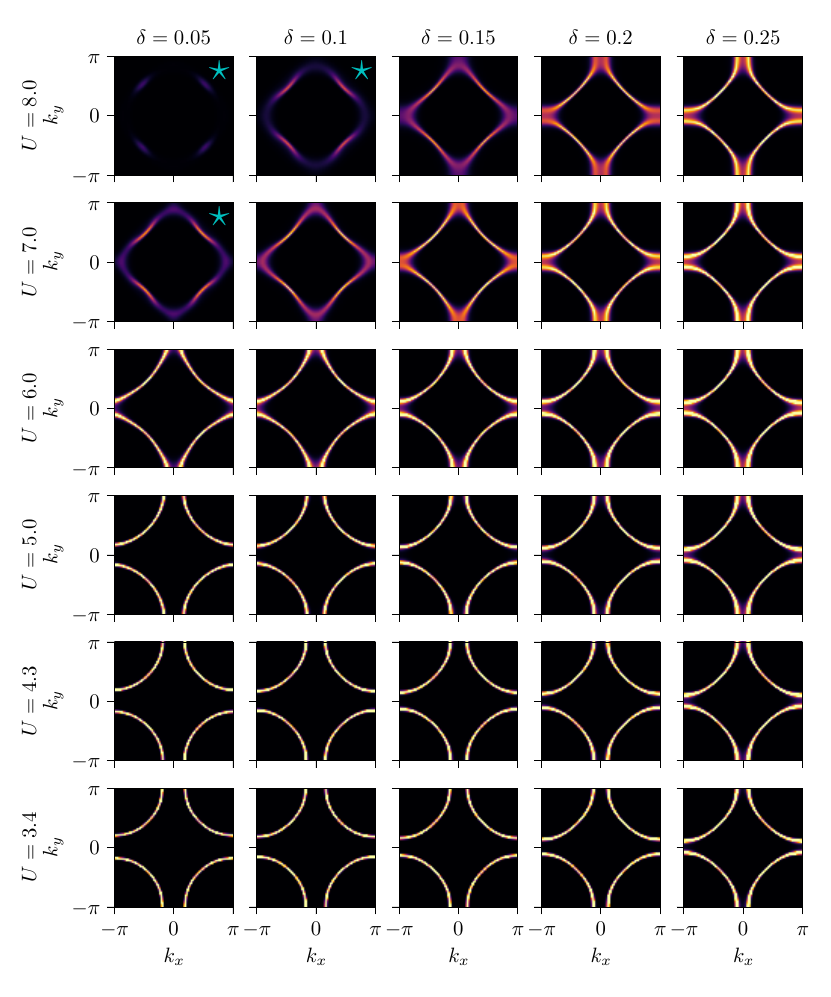}
     \includegraphics[scale=1.38,trim=0.6cm 0 0 0]{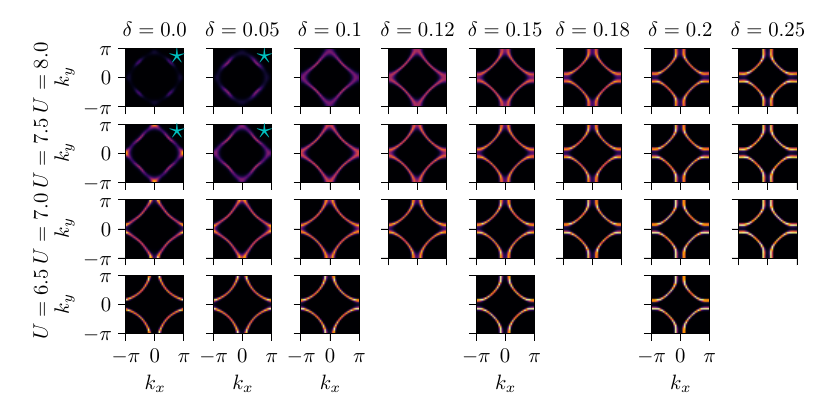}
    \caption{Integrated spectral function $-\frac{\beta}{\pi}G\left(\mathbf{k},\tau=\beta/2\right)$ at $\beta=100$ for 
        for $t'=-0.17,\ t''=0.13$, $U=8t$ \rep{(top)  and  $t'=-0.22,\ t''=0.14$, $U=8t$ (bottom). Cyan stars mark where $\chi^{-1}(i\omega = 0) < 0.01$ as an empirical proxy for the PG.}}
    \label{fig:PG}
\end{figure*}

In this Section, we supplement Figure~2 (a) of the main text by showing that the spectral function for the cyan dots actually displays a pseudogap.
Specifically, in Fig.~\ref{fig:PG} we plot
\begin{equation}
  -\frac{\beta}{\pi}G\left(\mathbf{k},\tau=\beta/2\right)=\frac{\beta}{2\pi}\int\text{d}\omega\frac{A\left(\mathbf{k},\omega\right)}{\cosh\left(\beta\omega/2\right)} \; ,
\end{equation}
which averages the spectral function  $A\left(\mathbf{k},\omega\right)$ over a frequency interval $\sim T$.
It has the advantage that it avoids the ill-conditioned analytical continuation. \rep{And good agreement with the empirically observed proxy $\chi^{-1}(i\omega = 0)\leq 0.01$ is observed.}

\begin{figure*}
    \centering
    \includegraphics[scale=0.9]{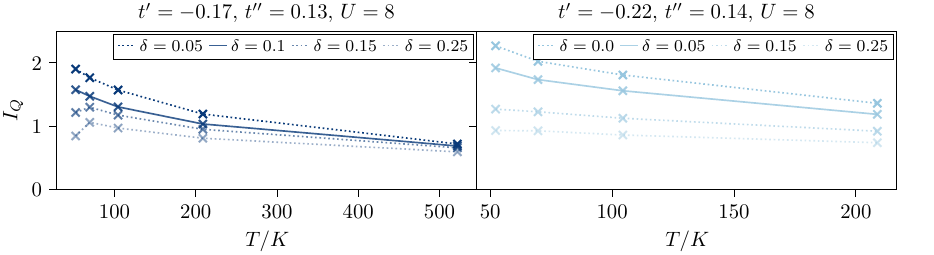}
    \caption{\rep{Temperature dependence of the quantum variance $I_Q$ for different dopings $\delta$. The solid line marks the value shown in the main text. (left) $t'=-0.17,\ t''=0.13$ at $U=8t$; (right)  $t'=-0.22,\ t''=0.14$ at $U=8t$.
      }}
    \label{fig:doping_dep_IQ}
\end{figure*}

\rep{In Fig.~\ref{fig:doping_dep_IQ}, we show how tuning the doping influences the quantum variance. Outside the PG regime, the growth of entanglement with lowered temperature is significantly reduced along with the overall value. In ladder D\textGamma A, the superconducting fluctuations do not influence the spin susceptibility. 
Such an effect is only included in the much more involved parquet D$\Gamma$A
\cite{Valli2015,Li2017}
Hence, in difference to the experimental results shown in the main text, the downturn observed at large values of $\delta$ for $t'=-0.17$ is not due to superconductivity. 
Moreover, the downturn of $I_Q$ is accompanied with a non-monotonicity in $\chi(i\omega=0)$. This requires further investigation beyond the scope of this work. }

\bibliography{references.bib}